\documentclass{emulateapj}
\usepackage{apjfonts}
\usepackage{lscape}
\input{epsf}

\slugcomment{To appear in the Astronomical Journal}

\shorttitle{Wide companions to nearby Hipparcos stars}
\shortauthors{Tokovinin \& L\'epine}

\begin{document}

\title{Wide companions to Hipparcos stars within 67 pc of the Sun}

\author{Andrei Tokovinin\altaffilmark{2}, \& S\'ebastien
  L\'epine\altaffilmark{2,3}}

\altaffiltext{1}{Cerro Tololo Inter-American Observatory, Casilla 603,
  La Serena, Chile, atokovinin@ctio.noao.edu}

\altaffiltext{2}{Department of Astrophysics, Division of Physical Sciences,
  American Museum of Natural History, Central Park West at 79th
  Street, New York, NY 10024, USA, lepine@amnh.org}

\altaffiltext{3}{Department of Physics, Graduate Center, City
  University of New York, 365 Fifth Avenue New York, NY 10016, USA}

\begin{abstract}
A  catalog of  common-proper-motion (CPM)  companions to  stars within
67\,pc of the Sun is constructed based on the SUPERBLINK proper-motion
survey.   It   contains  1392  CPM  pairs   with  angular  separations
$30\arcsec<\rho<1800\arcsec$, relative  proper motion between  the two
components less  than 25\,mas~yr$^{-1}$, magnitudes and  colors of the
secondaries  consistent  with  those  of  dwarfs  in  the  $(M_V,V-J)$
diagram.  In addition, we list 21 candidate white-dwarf CPM companions
with  separations under  300\arcsec,  about half  of  which should  be
physical.   We  estimate  a  0.31  fraction of  pairs  with  red-dwarf
companions to be physical systems  (about 425 objects), while the rest
(mostly  wide  pairs)  are  chance  alignments.   For  each  candidate
companion,  the probability  of a  physical association  is evaluated.
The distribution  of projected separations  $s$ of the  physical pairs
between  2\,kAU and  64\,kAU  follows $f(s)  \propto s^{-1.5}$,  which
decreases faster  than \"Opik's law.   We find that  Solar-mass dwarfs
have no less  than $4.4 \pm 0.3$\% companions  with separations larger
than 2\,kAU,  or $3.8 \pm 0.3$\%  per decade of  orbital separation in
the 2 to 16\,kAU range.  The  distribution of mass ratio of those wide
companions is approximately uniform in the $0.1<q<1.0$ range, although
we  observe  a dip  at  $q\simeq0.5$  which,  if confirmed,  could  be
evidence of bimodal distribution of companion masses. New physical CPM
companions to two exoplanet host stars are discovered.
\end{abstract}


\section{Introduction}
\label{sec:intro}

Very wide binaries play a special role in astrophysics. Low binding
energies make them sensitive probes of Galactic dynamical
environment. The very existence of wide binaries can constrain
properties of the dark matter \citep[e.g.][]{Weinberg87,Yoo04}. The
typical size of protostellar cores ($\sim$0.1\,pc) is comparable to the
separation of the widest binaries that survive against gravitational
perturbations in the Galaxy. Most stars form in clusters, in which
wide (and even relatively closer) binaries are susceptible to
disruption; the surviving wide pairs provide constraints on the
dynamical environments and processes at the epoch of star formation
\citep{Parker2009}. At larger scales of $\sim$1\,pc we also encounter
co-moving remnants of nearby clusters and associations
\citep[e.g.][]{Zuckerman01,Zuckerman04}. These groups of
stars, related by common origin, cannot be classified as physical
systems (binaries) because as far as we know, they do not move on
stable orbits around each other. The transition between {\it
 binaries} and {\it co-moving stars} is still an unsettled
issue. \citet{SO11} found pairs with projected separations of few pc
in the {\it Hipparcos}  catalog, many of which belong to  known moving
groups. \citet{Caballero10} argues that one multiple system with 1-pc
separation (member of  the  Castor moving group) could be  ``hard''
enough to remain bound.

Yet another use of wide binaries is in the calibration of luminosities
and abundances in the low-mass regime, referencing to their hotter
well-studied primary components. Parallaxes of the bright primaries
from, e.g., the Hipparcos catalog can be assigned to the secondaries
and provide accurate luminosity measurements \citep{GouldChaname04}. A
well-characterized primary is particularly useful to calibrate the
fundamental parameters of M dwarfs, whose complicated spectra
dominated by molecular absorption bands still resist detailed
atmospheric modeling. The metallicity scale of M dwarfs remains
largely dependent on metallicity measurements of FGK primaries in
common proper motion (CPM) pairs \citep{RojasAyala2012}. It is thus
important to assemble catalogs of common proper motion pairs where the
primaries are well-characterised {\it Hipparcos} FGK dwarfs, and the
secondaries are M dwarfs spanning a broad range of colors and
luminosities.

The main motivation of this study is to provide a reasonably complete
census of very wide companions to nearby Solar-type stars. Combined
with information on closer pairs also available for this sample, it
will lead to the un-biased multiplicity statistics much needed as a
benchmark for star-formation theories \citep{Bate09}. The recent
study by \citet{Raghavan10} furnished such statistics for 454
stars within 25\,pc from the Sun, but a much larger sample is needed  to
study triple and higher-order hierarchies because they comprise only
$\sim$10\% of all objects. Most solar-type exo-planet hosts are also beyond the
25\,pc horizon of previous multiplicity surveys. 
It is  important to search for  and identify wide  companions to known
exoplanet hosts, and companions to  other nearby stars which are still
being  observed in  search of  exo-planets, in  order  to characterize
planet populations in binary systems \citep{Roell12}.

Primary targets for our search for CPM companions are all stars with
parallax $\pi_{\rm HIP} > 15$\,mas selected from the {\it Hipparcos}-2
catalog \citep[][hereafter HIP2]{HIP2}. The {\it Hipparcos} catalog is complete to
67\,pc for dwarfs of mid-G spectral type or brighter, thus providing a
clean distance-limited sample for multiplicity statistics. At the same
time, these nearby stars move fast enough to discriminate CPM
companions against background stars and are close enough to detect
intrinsically faint companions. Little is to be gained by extending
the distance limit further.

Existing data on CPM binaries is in large part based on the Luyten visual
searches for CPM binaries with $\mu>0.1\arcsec$~yr$^{-1}$ \citep[LDS,
][]{LDS}. A recent study has repeated the visual search, by blinking
the Digitized Sky Survey fields around exo-planet host stars
\citep{Raghavan06} and stars within 25\,pc
\citep{Raghavan10}. A more systematic search for companions to
{\it Hipparcos} stars was conducted by \citet[][hereafter LB2007]{LB07},
who used the LSPM-north catalog of
stars with proper motions $\mu>0.15\arcsec$ yr$^{-1}$ \citep{LS05};
the paper also gives a good overview of prior work. Detecting CPM
binaries in large databases was also demonstrated by
\citet{Chaname05}, \citet{Makarov08}, and \citet{Dhital10}, for
systems with fainter primaries.

In this paper, we expand  on the search performed by LB2007, and use
the full SUPERBLINK all-sky proper motion catalog, which is an
extension of the published LSPM-north catalog. The method for
selecting CPM companion candidates is presented in
Section~\ref{sec:cat}. Statistics of all companions are discussed in
Section~\ref{sec:comp}. Then we concentrate in Section~\ref{sec:stat}
on the statistics of  wide physical binaries, namely the distributions
of  projected  separations  and   mass  ratios.  Our  conclusions  and
discussion of results are presented in Section~\ref{sec:sum}.

\section{Identification of common proper motion companions}
\label{sec:cat}

\begin{figure}[t]
\epsscale{2.3}
\plotone{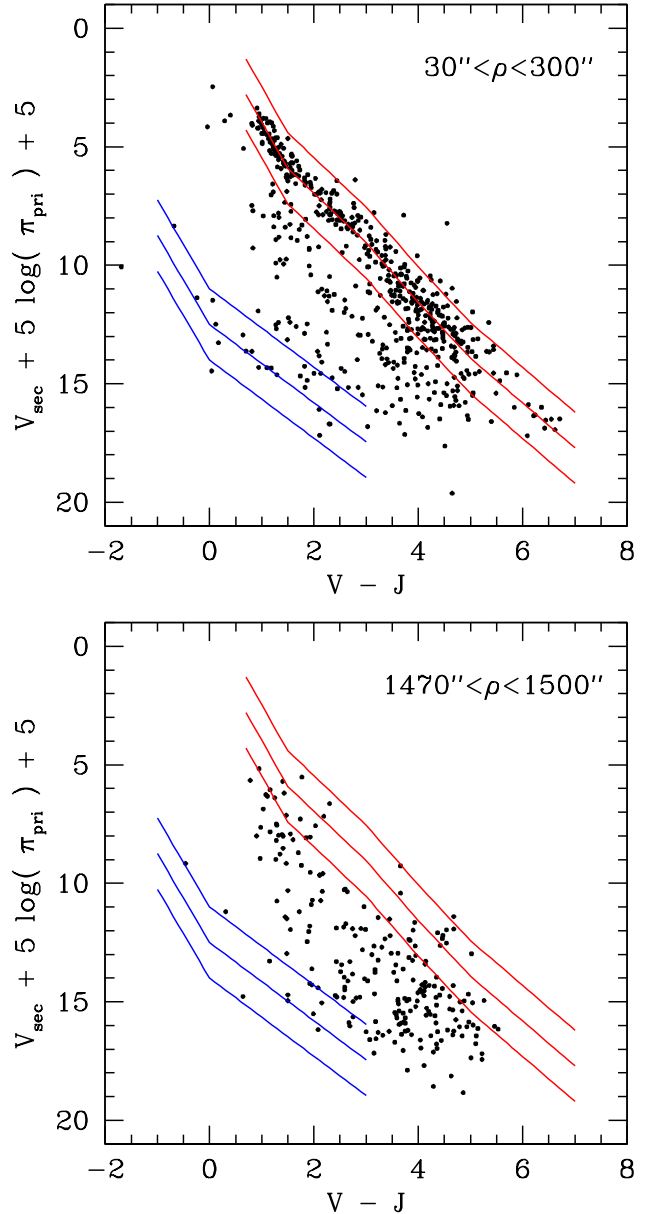}
\caption{Color magnitude diagram of the common proper motion
  secondaries, assuming they have the same parallax as the ({\it Hipparcos})
  primaries. Physical companions are expected to fall either close to
  the main sequence (red) or the white dwarf sequence
  (blue); a scatter of $\pm$1.5 mag is adopted for the
  selection (shown). Stars falling outside the bounds are assumed to
  be chance alignments of unrelated stars. Top: diagram for the 608
  pairs with separations $30\arcsec<\rho<300\arcsec$. Bottom: diagram
  for the 240 pairs with separations $1470\arcsec<\rho<1500\arcsec$,
  which cover a similar search area. Pairs in the first sample are
  more numerous and tend to concentrate within the expected main
  sequence and white dwarf sequence, suggesting that the $\approx$400
  extra pairs in the first distribution are physically
  related.\label{fig:cmd}}
\end{figure}

\subsection{SUPERBLINK selection}

The  SUPERBLINK survey  is based  on  an automated  search for  moving
sources  from multi-epoch images  of the  Digitized Sky  Survey (DSS),
notably  scans from  the first  and second  epoch Palomar  Sky Surveys
(POSS-I, POSS-II). Compared to  LSPM-north, the detection threshold in
the  full  SUPERBLINK catalog  is  significantly lower,  $\mu>40$\,mas
yr$^{-1}$ for stars north  of declination $-20^\circ$, and extends the
survey  to a  proper  motion limit  $\mu>150$\,mas  yr$^{-1}$ for  all
declinations south  of $-20^\circ$. The  survey is still  in progress;
more   information   can   be   found   in   the   recent   paper   by
\citet{LG11}.  SUPERBLINK is  at least  90\% complete  down  to visual
magnitude  $V=19$,  and has  a  very  low  level of  false  detections
($<0.1\%$)  thanks  to  stringent  quality controls  including  visual
confirmation of all  targets.  The $V$ magnitudes  of faint stars
  are estimated  from the DSS photographic photometry reported in
  \citet{Monet03}, using the procedure described in \citet{LS05}.

We select as possible primaries of CPM pairs all
stars from HIP2 with parallax greater
than 15\,mas and proper motion larger than the SUPERBLINK limit ($\mu
> 40$\,mas yr$^{-1}$ for stars north  of declination $-20^\circ$ and  $\mu >
150$\,mas yr$^{-1}$ otherwise). Potential CPM companions to these
stars were selected from SUPERBLINK by the following three criteria.
\begin{enumerate}
\item
Separation of the  companion $\rho$  should be in the  interval $30\arcsec \le
\rho \le 1800\arcsec$. Closer companions are affected by the oreols of
primary  stars  in  DSS  photographic  images,  introducing  selection
against  faint   companions.   Most  pairs   wider  than  $300\arcsec$
(projected  separation  20\,kAU   at  67\,pc)  are  chance  alignments
(optical), but including them in  the catalog helps us to evaluate the
proportion of true (physical) wide binaries.

\item
Proper  motion  difference with  the  primary  component $\Delta  \mu$
should be less than 25\,mas~yr$^{-1}$. Typical proper motion errors in
SUPERBLINK  are 8\,mas~yr$^{-1}$,  so  this cutoff  allows for  larger
errors or some additional proper motion difference caused by motion in
hierarchical sub-systems.

\item
Color and magnitude of the secondary should be consistent with a main sequence
star or a white dwarf at the same distance as the primary. 
This criterion effectively requires that the
photometric distance of the secondary as estimated from ($V$,$V-J$) be
consistent with the parallax distance of the alleged {\it Hipparcos}
primary. 
\end{enumerate}
The first two criteria identify 130\,882 possible pairings, the vast
majority of which are simple chance alignments. For the third
criterion, we first calculate the absolute magnitudes of the CPM
companions ($[M_V]_{\rm SEC}$) assuming that the secondary is at the
same distance as the {\it Hipparcos} primary:
\begin{equation}
[M_V]_{\rm SEC} = V_{\rm SEC} + 5 \log( \pi_{\rm PRI} ) + 5 ,
\end{equation}
where $\pi_{\rm PRI}$ is the {\it Hipparcos} parallax of the
primary. We then use the ($M_V$,$V-J$) color magnitude relationship as
calibrated in \citep{Lepine05} for main sequence (MS) stars  in the
SUPERBLINK survey. These define photometric absolute magnitudes
$[M_V]_{\rm MS}$ estimated from $V-J$:
\begin{displaymath}
[M_V]_{\rm MS} = 0.08 + 3.89 (V-J) \ \ \ \ \ \ \ [V-J < 1.5]
\end{displaymath}
\begin{displaymath}
[M_V]_{\rm MS} = 2.78 + 2.09 (V-J) \ \ \ \ \ \ \ [1.5 < V-J < 3.0]
\end{displaymath}
\begin{displaymath}
[M_V]_{\rm MS} = 1.49 + 2.52 (V-J) \ \ \ \ \ \ \ [3.0 < V-J < 4.0]
\end{displaymath}
\begin{displaymath}
[M_V]_{\rm MS} = 2.17 + 2.35 (V-J) \ \ \ \ \ \ \ [4.0 < V-J < 5.0]
\end{displaymath}
\begin{equation}
[M_V]_{\rm MS} = 4.47 + 1.89 (V-J) \ \ \ \ \ \ \ [5.0 < V-J < 9.0].
\label{eq:MV-VJ}
\end{equation}
In addition, we introduce the following simple color-magnitude
relationship which we use to define absolute magnitudes for white
dwarfs (WD) based on $V-J$:
\begin{displaymath}
[M_V]_{\rm WD} = 12.5 + 3.75 (V-J) \ \ \ \ \ \ \ [-1 < V-J < 0]
\end{displaymath}
\begin{equation}
[M_V]_{\rm WD} = 12.5 + 1.89 (V-J) \ \ \ \ \ \ \ [0 < V-J < 3].
\label{eq:WD}
\end{equation}
We then select only pairs  whose $[M_V]_{\rm SEC}$ agree to within 1.5
magnitudes  of either  photometrically  estimated absolute  magnitudes
$[M_V]_{\rm MS}$ or $[M_V]_{\rm WD}$. For 124 companions without 2MASS
photometry   the  $V-J$   color  is   estimated  from   the  $(b,r,i)$
photographic magnitudes as described by \citet{LS05}.

The color-magnitude selection  is illustrated in Figure~\ref{fig:cmd}.
The top panel  shows $[M_V]_{\rm SEC}$ as a function  of $V-J$ for all
608 pairs  with angular separations  $30\arcsec<\rho<300\arcsec$, with
the selection range  for MS stars in red, and  the selection range for
WDs shown in blue. The  bottom panel in Figure~\ref{fig:cmd} shows the
same distribution  but for 240  pairs with angular separations  in the
$1470\arcsec<\rho<1500\arcsec$ range.  Note  that the two ranges cover
the same  area on  the sky,  so if all  CPM pairs  were the  result of
chance alignments,  one would expect  the two samples to  have similar
numbers. Instead we find an excess of 368 pairs at closer separations.
In addition, the color-magnitude distribution is very different in the
two  samples.   Most  pairs with  $1470\arcsec<\rho<1500\arcsec$  fall
outside of the  MS or WD bands, while the majority  of the close pairs
have colors  consistent with true companions.   This demonstrates that
$\sim400$  CPM  pairs  in  our catalog  are  physical,  binary/multiple
systems.

\begin{figure}[t]
\epsscale{1.15}
\plotone{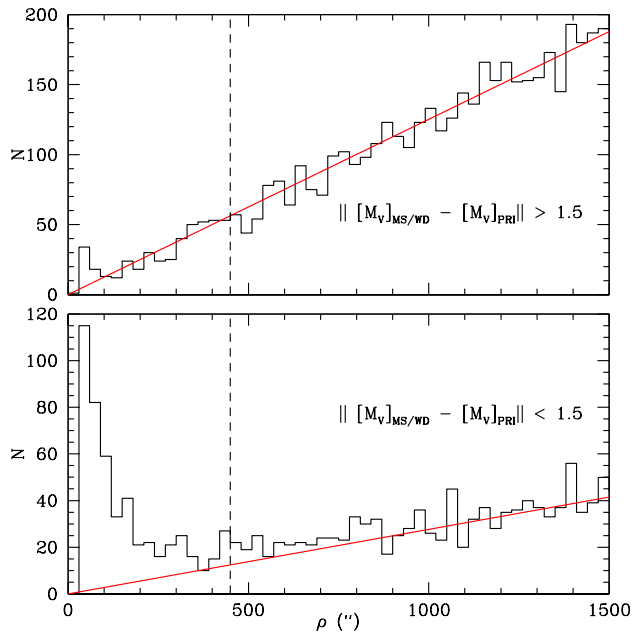}
\caption{Histogram of the number of pairs found as a function of
  angular separation $\rho$. Top: distribution for pairs assumed to be
  chance alignments, based on the inconsistency in the parallax of the
  primary and photometry of the secondary, i.e. stars falling outside
  of the MS or WD sequences (see Fig.~\ref{fig:cmd}). Bottom:
  distribution for pairs in which the secondary would have ($V,V-J$)
  consistent with a main sequence star or white dwarf. In either case,
  the   red line shows the distribution expected from chance
  alignments (linear increase with $\rho$). Pairs in which the
  color/magnitude of the secondary is   consistent with a physical
  companion show a net excess  at $\rho<200\arcsec$. 
\label{fig:disthist}}
\end{figure}

The color-magnitude selection reduces  the catalog to only 2\,078 pairs
with  consistent colors.   A histogram  of the  number of  pairs  as a
function   of   their   angular   separation  $\rho$   is   shown   in
Figure~\ref{fig:disthist}. The distribution  for pairs with consistent
parallax/photometric  distances  (bottom  panel)  is compared  to  the
distribution  of pairs  where the  primary and  secondary do  not have
colors consistent  with physical association (top  panel).  The latter
distribution shows the trend  expected from chance alignments, i.e. a
linear increase with  $\rho$ proportional to the increase  in the area
of  sky; this is  modeled by  the red  line.  The distribution of
stars with consistent MS or WD color-magnitude, however, shows  a
marked excess of pairs  with $\rho<200\arcsec$, which again points  to
the  existence of physical  pairs. At  larger angular separations
$\rho>450\arcsec$,  the  distribution  shows  the  linear increase
expected from  chance alignments;  the red  line  shows the predicted
trend based on  the number of pairs with $\rho>1000\arcsec$. One
notices that  while a  significant number  of physical  pairs are
undoubtedly detected at  $\rho<450\arcsec$, this sub-sample
still suffers from contamination from chance alignments.

We note that some companions share common proper motion with more than one
{\it Hipparcos} star. Likewise some {\it Hipparcos} stars have several CPM
companions. Some are genuine multiple systems (triples), but in
most cases they are more likely to be the result of chance alignments,
e.g. the alignment of a foreground/background proper motion star with
a true physical pair. In any case, all possible pairings are retained
in our list.

\subsection{Treatment of the White Dwarf CPM Companions}

Magnitudes and colors  of 386 companions place them in  the WD band in
Figure~\ref{fig:cmd}.   They  are found  in  smaller  numbers than  MS
candidates,    and     are    significantly    more     affected    by
contamination. Indeed, only 21 pairs have $\rho <300\arcsec$, and some
of those are expected to be chance alignments. We evaluated the number
of  physical WD  companions  within some  separation  $\rho$ by  their
excess over the  linear distribution, as in Figure~\ref{fig:disthist},
and found that the number  of likely physical pairs with WD companions
does not  increase beyond $\rho  > 200\arcsec$, reaching an  excess of
around 10 pairs.

Since the vast majority of the stars falling in the WD region are thus
chance alignments, we  removed form the catalog all  companions in the
WD  band with $\rho  > 300\arcsec$,  keeping only  the 21  most likely
pairs.

\subsection{Exclusion of nearby cluster stars}
\label{sec:cluster}

The selection of CPM pairs is sensitive to faint members of the Hyades
and Pleiades clusters, whose members have proper motions within the
$\mu>$ 40 mas yr$^{-1}$ limit of the  SUPERBLINK catalog, and are thus
generally detected. Some nearby young  associations and moving  groups
also  have members detected in SUPERBLINK \citep{Schlieder2012}.  Faint
cluster  stars share  a proper motion  with  the brighter members,
which are also {\it Hipparcos} sources. Cluster stars have photometric
distances consistent  with the  parallax of  the brighter members, so
they will generally  meet our selection criteria  for CPM pairs with
colors and magnitudes consistent with physical companions.

We  wish, however,  to exclude  those  pairings because  they are  not
physical binaries. We thus eliminate 311 pairs where the primary stars
are known members  of clusters and associations according  to the XHIP
catalog \citep{XHIP}. Of those, 230 belong to the Hyades.  This leaves
1392 CPM companions with MS colors and 21 companions with WD colors in
the catalog.

\subsection{Multiple stars}
\label{sec:mult}

In the catalog compilation and analysis presented below we assume
implicitly that both the primary and wide secondary companion are
themselves single stars, i.e. there are no unresolved companions. In
reality a substantial but still poorly known fraction of the
components in wide systems are themselves close binaries
\citep[e.g.][]{Makarov08}.  Internal orbital
motion can  contribute additional components to the proper motions
and increase $\Delta \mu$ with the primary. Analysis of astrometric
binaries among  FG dwarfs within  67\,pc listed by \citet{MK05} shows
that for 86\% of them the difference between long-term and {\it Hipparcos}
proper motion is less than 25\,mas yr$^{-1}$. Therefore, our catalog may be
slightly biased against triple systems.

To  verify this, we  simulated a  realistic population  of sub-systems
around primary targets with periods from 1\,yr to 1000\,yr susceptible
to create $\Delta  \mu$ by photo-center motion during  the 3.2-yr long
{\it Hipparcos}  mission.  Maximum effect is produced  at periods from
4\,yr to  40\,yr, but the mean  and median $\Delta \mu$  do not exceed
6\,mas~yr$^{-1}$ at  50\,pc distance; in  90\% of cases $\Delta  \mu <
12$\,mas~yr$^{-1}$ for  all periods.  According  to these simulations,
the  fraction  of  multiple   systems  rejected  by  the  $\Delta  \mu
<25$\,mas~yr$^{-1}$ threshold should be significantly smaller than the
14\% implied by the paper  of \citet{MK05}. The discrepancy is alleviated
by including in the simulation  some massive companions that are close
binaries and cause larger $\Delta \mu$ \citep{Tok12}.

Multiple  systems  can also  be  missed  because of inaccurate or
distorted   photometry. Visual magnitudes in the SUPERBLINK catalog
are largely based on photographic measurements, and sometimes have
large errors, especially if the star is blended with another source,
such as in the vicinity of a very bright primary or in a dense
field. The   composite  magnitudes   of  unresolved sub-systems also
bias photometric distance estimates which might yield to rejection
based on color and magnitude \--- although the 1.5 magnitude selection
range (see Section~ 2.1) mitigates this effect. In any case, this may
be another non-negligible, but  poorly quantified, incompleteness
factor for hierarchical systems with CPM companions.

\section{Catalog description}
\label{sec:table}

The complete catalog of 1413 CPM  pairs is listed in Table~1, the full
version  of which  is available  electronically. The  presentation has
some similarity with LB2007.  The first columns contain information on
the primary  target: (1) {\it  Hipparcos} number HIP$_{\rm  PRI}$, (2)
SUPERBLINK  identification, (3,4) equatorial  ICRS coordinates  at the
2000.0  epoch  in degrees,  (5,6)  proper motions  in  RA  and Dec  in
arcsec~yr$^{-1}$, (7)  $V$ magnitudes derived from the  DSS for faint
stars or taken from {\it Tycho-2} for bright ones, (8-11) $V-J$ color and $J,
H,  K  $  magnitudes  from  the 2MASS  All-Sky  Point  Source  Catalog
\citep{2MASS}.  Then follow the parallax (12), its error (13), and the
parallax code (14), namely  PLX1 for trigonometric parallax from  HIP2.
  The  following  columns list  (15)  separation
$\rho$ in  arcseconds, (16) proper motion difference  with the primary
target $\Delta  \mu$ in mas~yr$^{-1}$, and  (17) estimated probability
of         physical         association         $P_{\rm         phys}$
(Section~\ref{sec:pphys}). Companions  in the WD band  (see Section~2.2)
are distinguished by $P_{\rm phys}=-1$ and listed at the end.

The remaining columns  of Table~1 contain data on  the CPM companions.
Columns (18)  to (31) give  same information for companion  as columns
(1) to  (14) for the primary.   The HIP$_{\rm SEC}$ number  in (15) is
999999 if  not available. Similarly, missing magnitudes  are listed as
99.  Photometric parallaxes in column (29)  estimated from the
$M_V, V-J$ relation have code PHOT in column (31), their errors are set to zero.
 Parallax of some companions is taken from the literature
and      marked      by       the      following      codes:   
PLX2 \citep{vanAltena95}\footnote{\url{http://vizier.cfa.harvard.edu/viz-bin/VizieR?-source=I/238A}},
PLX3 from the NSTARS database\footnote{\url{http://nstars.nau.edu/nau\_nstars/index.htm}}, 
and               PLX4               \citep{Myers02}\footnote{
  \url{http://cdsarc.u-strasbg.fr/viz-bin/Cat?V/109}}.          Trigonometric
parallaxes  of  two  WDs    \citep{PLX5}  have  code  PLX5,  the
photometric parallaxes of WDs are calculated from (\ref{eq:WD}). 


We emphasize again that this  catalog contains all CPM pairs identified
using the  selection method  described in \S2,  and includes  not only
physical   binaries  but   also   a  significant   number  of   chance
alignments.  One should therefore  {\em not}  assume that  the catalog
lists  true companions  of the  {\it Hipparcos} stars, and the user
should keep an eye on column 17, which provides a probability for the
pair to be physical.  We describe  below our statistical  method to
assign  probabilities of physical association for  all the CPM  pairs.

The 21 candidate WD  companions are highlighted in Table~\ref{tab:wd},
extracted from  the main  catalog. The {\it  Hipparcos} number  of the
primary component HIP$_{\rm PRI}$ is listed in column (1), followed by
the equatorial coordinates of  the companion for 2000.0 in hexadecimal
format in  columns (2) and (3). Subsequent  columns contain separation
(4), $\Delta \mu$ (5), proper motion of the primary component (6), its
parallax (7),  $V_{\rm SEC}$  magnitude of the  companion (8),  and its
$V-J$  color or  its estimate  (9).  The  last column  (10)  gives the
identifications of 5 previously known  WDs found in SIMBAD.  Another 6
promising WD candidates are marked as wd?  in this column on the basis
of their  large proper motion  or small separation.  The  companion to
HIP 118010  is also  listed in LB2007.  HIP~60081 hosts a  planet with
48\,d period \citep{Schneider}.

\section{Statistical properties of the catalog}
\label{sec:comp}

Half  of our  targets have  $\mu  < 87$\,mas yr$^{-1}$  and 3/4  have
$\mu  < 150$\,mas yr$^{-1}$.  Therefore  the  extension of the
SUPERBLINK proper motion limit  to 40\,mas yr$^{-1}$ is critical  for
this work.  On the other hand, the catalog is still  incomplete at
$\delta < - 20^\circ$.  The  distribution of companions on the sky is
uniform, but with fewer stars south of $\delta = -20^\circ$ due  to
the larger proper motion cutoff.  Interestingly, there  is no obvious
concentration towards the Galactic plane. Although the density of
background  stars strongly depends  on the Galactic  latitude, they
mostly have small $\mu$ and thus are not included as companion
candidates \citep[see][]{LG11}.

Selection of primary targets with relatively large proper motion
naturally favors  nearby stars. Median parallax of the catalog entries
is 18.1\,mas. The number of FG dwarfs within  distance $d$  is
proportional to  $d^{2.1}$  in our  catalog, whereas it is close to
$d^3$ in {\it Hipparcos}; the difference is caused by the proper
motion selection.  In this Section we do  not consider  the 21  WD
candidate companions and analyze only the 1392 MS companions.

The  parallax  distributions for  close  and  wide  companions in  our
catalog are  slightly different; the median parallax  is 22.1\,mas and
17.1\,mas   for   $\rho    <   300\arcsec$   and   $\rho>1200\arcsec$,
respectively.   Distant stars  have, on  average, a  smaller proper
motion  and a larger  number  of  optical  companions,  producing
this  correlation between $\pi_{\rm HIP}$  and $\rho$.

\subsection{Proper motion difference}
\label{sec:dmu}

\begin{figure}[t]
\epsscale{1.1}
\plotone{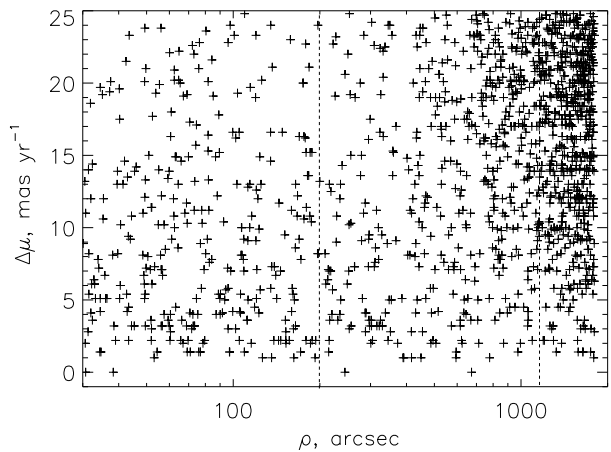}
\caption{Proper motion difference $\Delta \mu$ vs. angular separation
  $\rho$ for the 1\,392 common proper motion pairs with MS
  secondaries in our catalog. Pairs with short separations are
  dominated by physical systems, whereas pairs with large
  separations are mainly chance alignments. Vertical dashed lines mark separations of 200\arcsec and 1200\arcsec.
\label{fig:dmu-sep}}
\end{figure}

\begin{figure}[t]
\epsscale{1.1}
\plotone{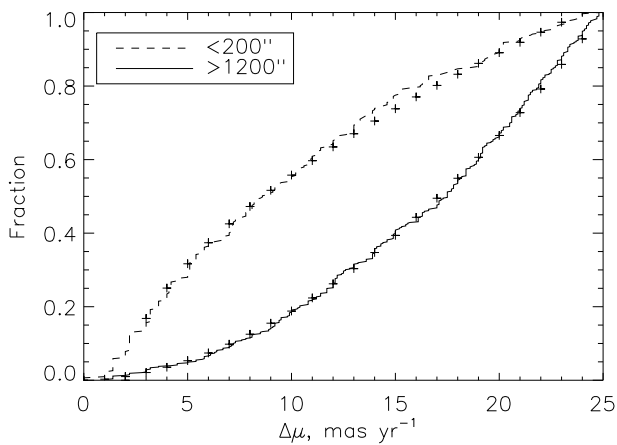}
\caption{Cumulative distributions  of $\Delta \mu$ for  close ($\rho <
  200\arcsec$, $N=315$,  dashed line) and  wide ($\rho >
  1200\arcsec$, $N=552$,   full line) CPM pairs. The crosses show
  analytical fits. Close pairs show a distribution which decreases in
  $\Delta  \mu$, consistent with physical systems, while wider pairs
  show one that is increasing in $\Delta \mu$, consistent with chance
  alignments. \label{fig:dmu}}
\end{figure}

Figure~\ref{fig:dmu-sep}   plots proper motion difference $\Delta \mu$
for the pairs with main sequence secondaries against their angular
separation $\rho$. The distribution reveals  statistical   differences
between physical (close) and optical  (wide) companions. Companions
with $\rho <200\arcsec$ have uniform distribution of  $\log \rho$ and
mostly $\Delta \mu <15$\,mas yr$^{-1}$.  In  contrast,  wide
companions  with $\rho>1200\arcsec$   
tend  to  have larger  $\Delta \mu$. The difference of the  $\Delta
\mu$ statistics between these two groups  is  illustrated in
Figure~\ref{fig:dmu}: while  56\% of  close companions  have $\Delta  \mu
<10$\,mas~yr$^{-1}$,  this cutoff  contains only 19\% of wide
companions.  Cumulative distribution for wide companions can be
fitted by a power law:
\begin{equation}
F_{\rm opt}(\Delta \mu) \approx  (\Delta \mu/25)^{1.78} .
\label{eq:dmu1}
\end{equation}
If the proper motion vectors  of background  stars were  distributed
uniformly, one would expect a  quadratic law (LB2007), but the  actual
power index is slightly  less than two; this is because nearby stars
have systemic motions relative to the Sun and their proper motions
thus show local correlations on the sky. The analytical  model for
close (physical) companions is:
\begin{equation}
F_{\rm phys}(\Delta \mu) \approx [(\Delta \mu - 2)/22]^{0.57}.
\label{eq:dmu2}
\end{equation}
These  two  distributions overlap  substantially,  preventing us  from
using  a smaller cutoff  in $\Delta  \mu$.  Both  distributions extend
beyond  the  25\,mas  yr$^{-1}$  cutoff (the  cumulative  curves  have
non-zero slope  near the cutoff),  suggesting that the  catalog misses
some physical  companions with  $\Delta \mu >25$\,mas  yr$^{-1}$, e.g.
wide triples  with motion  in inner sub-systems.   Physical companions
with  $\Delta  \mu  >25$\,mas   yr$^{-1}$  are  indeed  identified  in
LB2007. However,  the large number  of chance alignments  with $\Delta
\mu  >25$\,mas yr$^{-1}$  places very  low probabilities  on  any pair
being physical  and would require significant  follow-up resources for
triage  and  confirmation.  For  now,  we prefer  to  adhere  to  this
(somewhat arbitrary) upper limit in $\Delta \mu$.

\subsection{Magnitudes and colors}

\begin{figure}[t]
\epsscale{1.1}
\plotone{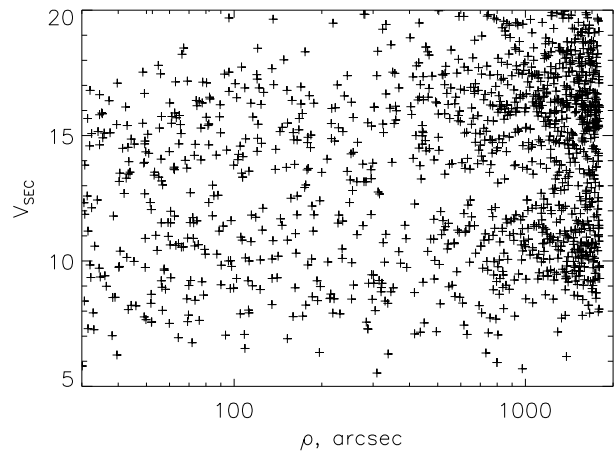}
\caption{Visual magnitude of the secondaries $V_{\rm SEC}$ as a function
  of the angular separation $\rho$ of the CPM pair. The secondaries in
  close ($\rho<100\arcsec$) pairs, which are largely physical systems,
  have a relatively uniform magnitude distribution. Secondaries in
  wider (and mostly change alignment) pairs show a magnitude gap at
  $V_{\rm SEC} \approx 13$ which can be explained by a combination of selection
  effects.\label{fig:V2-sep}}
\end{figure}

In our catalog  of CPM binaries, secondaries in wide  pairs tend to be
fainter   than  those   in  closer   pairs  (Figure~\ref{fig:V2-sep}).
Detection of faint  close companions is affected by  the oreols around
bright primary  targets (LB2007), but at $\rho>30\arcsec$  it seems to
be complete at  least down to $V_{\rm SEC} =17^m$. 

There appears to  be a  deficiency of  companions with  $V_{\rm SEC}$ from
$12^m$ to  $14^m$. Bright candidates  with $V_{\rm SEC} <12$  were selected
from  the   {\it  Tycho-2},  while fainter  candidates are
identified directly from the  DSS with SUPERBLINK. There is,
however, an overlap of a couple magnitudes between the {\it Tycho-2}
catalog and SUPERBLINK identifications. \citet{LS05} argued that
the merged {\it Tycho-2} + SUPERBLINK catalog should be 98\% complete in
this overlapping range of magnitudes, based on the high rate of
recovery by SUPERBLINK of the faintest of the {\it Tycho-2} stars.
However, the $V_{\rm SEC}$ magnitudes of faint companions are still derived
from the DSS scans of photographic plates, with estimated  errors of
$\pm  0.5^m$ or  larger. At the bright end, saturation of the stars on
the plates may lead to systematic errors, and a slight
over-estimation  of $V$ could create the discontinuity of the $V_{\rm SEC}$
distribution seen in Figure~\ref{fig:V2-sep}.

There is, however, an alternative explanation. One can see from
Figure~\ref{fig:V2-sep} that the gap is most apparent in the very wide
secondaries,  which are largely dominated with chance alignments,  and
much  less  obvious in  the close  ($\rho<100\arcsec$) pairs.   While
any  field star  could potentially  be selected  as CPM secondary
based  on   proper  motion   alone,   our  color-magnitude requirement
eliminates  most  of   the  foreground   and  background contaminants,
because their colors  and magnitude  would not  fit the expected MS at
the distance of the star. Exceptions  to this rule are background
giant  and  subgiant  stars,  which  generally  cannot  be
distinguished   from  foreground   MS   stars  based   on  color   and
magnitude. Effectively,  this causes  an excess of  (chance alignment)
wide  CPM  secondaries  in  the $7<V_{\rm SEC}<12$  magnitude  range;  fainter
subgiants  are not  found in  the  SUPERBLINK catalog  because of  the
distance  bias  introduced  by   the  high  proper  motion  cut.  This
over-selection  of bright giants/subgiants  conspires with  the rising
luminosity function  of field  stars to create  the magnitude  gap for
wide companions.

\begin{figure}[t]
\epsscale{1.1}
\plotone{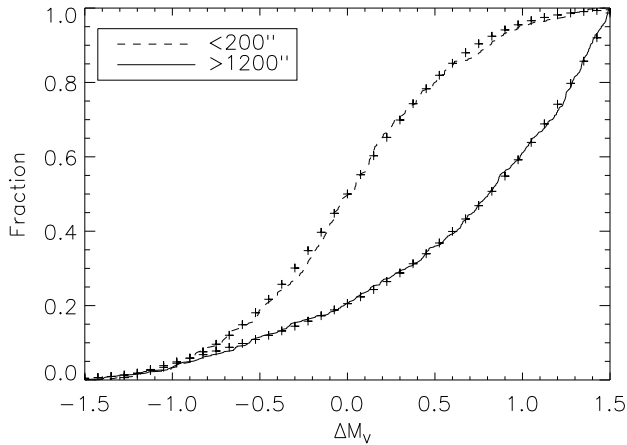}
\caption{Cumulative  distributions of  $\Delta M_V$  for  close ($\rho
  <200\arcsec$, $N=315$,  dashed line) and  wide ($\rho >1200\arcsec$,
  $N=552$,  full line)  companions. Analytical  models are  plotted as
  crosses.
 \label{fig:dmag}}
\end{figure}

A color-magnitude diagram  (CMD) of  the companion candidates is  plotted in
Figure~\ref{fig:cmd}.   There is no  clear separation  between optical
and  physical  companions in  this  CMD,  but  the statistics  of  the
deviation  from  the  nominal  MS  $\Delta M_V  =  [M_V]_{\rm  SEC}  -
[M_V]_{\rm  MS}$ (see  Section  2.1) is  different  between those  two
groups.  Most  close (physical) companions  have $\Delta M_V  \sim 0$,
while  wide companions  tend to  be fainter  and mostly  have positive
$\Delta M_V$.   Cumulative distributions of $\Delta  M_V$ are compared
in  Figure~\ref{fig:dmag}. The  cumulative distribution of $\Delta
M_V$ for wide companions is  approximated by a cubic polynomial  in
$\Delta M_V$ with  coefficients $(0.2052, 0.2342, 0.1210,
0.0468)$. The $\Delta  M_V$  distribution of  close companions is
however normal with a dispersion of $0.575^m$ and zero average. Our
cutoff of $\pm 1.5^m$ ($2.6\sigma$) rejects  only  1\%  of   this
normal distribution. This indicates that our color-magnitude selection
misses very few MS companions. A reduction of the cutoff would likely
reduce the contamination by optical pairs, but would result in a lower
completeness for physical systems.

\subsection{Distribution of angular separations}

\begin{figure}[t]
\epsscale{1.1}
\plotone{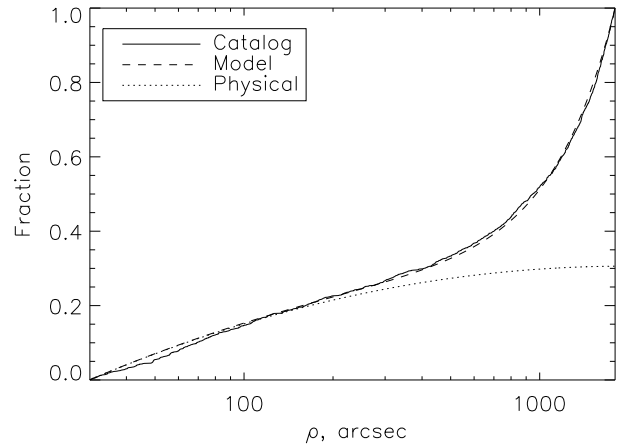}
\caption{Cumulative  distribution of separations  (full line)  and its
  model  (dashed  line).   The  dotted line  corresponds  to  physical
  companions.
\label{fig:model} 
}
\end{figure}

We expect  that the  number of {\em optical}  (i.e. chance alignment)
companions closer  than $\rho$ should be proportional to $\rho^2$.
Physical companions, on the other hand, are expected to be distributed
almost uniformly in $\log \rho$ \citep[\"Opik's   law,][]{Opik}. This
allows us to model  the cumulative distribution of separations as a
sum of physical and optical populations,
\begin{equation}
F(\rho) = ax + bx^2 + c(\rho/\rho_2)^2,
\label{eq:model}
\end{equation}
where:
\begin{displaymath}
x = \log(\rho/\rho_1) / \log (\rho_2/\rho_1).  
\end{displaymath}
Here $\rho_1 =  30\arcsec$ and $\rho_2 = 1800\arcsec$  are the
separation limits of  the  catalog. The  variable   $x$ maps this
interval logarithmically into $(0,1)$, and the parameters $a$ and $b$
describe the distribution of physical companions linear in $x$
(instead of uniform distribution we allow  some slope $b$), while $c = 1
- a -  b$ is  the fraction of  optical companions.  This  simple model
with only  two free  parameters fits well  the observed  distribution
of separations   in  Figure~\ref{fig:model}   with   $a=0.600$  and
$b= -0.294$. The  estimated fraction of optical companions  in our
catalog is therefore $c = 0.69$, so it appears to contain about 425
physical wide pairs. The same model predicts that  92.7\% of
companions with $\rho < 300\arcsec$ are physical.

\subsection{Probability of physical association}
\label{sec:pphys}


We  have  modeled the  distributions  of  two astrometric  parameters,
angular separation  $\rho$ and proper motion  difference $\Delta \mu$,
by   fitting  analytical   formulae  to   the   cumulative  histograms
(eqs.    \ref{eq:dmu1}--\ref{eq:model}).    We   also    modeled   the
distribution  of the deviation  from the  main sequence  $\Delta M_V$.
Probability density functions for  physical (close) and optical (wide)
companions are readily obtained by differentiating these formulae.  We
assume  here  statistical independence  of  variables $\rho$,  $\Delta
\mu$, and $\Delta M_V$, which  is not quite true owing to correlations
between  these   parameters,  but  still   a  reasonable  approximation.
Neglecting the correlations, the three-dimensional probability density
functions for  physical and optical  companions are products  of their
one-dimensional  distributions,  $f(\rho, \Delta  \mu,  \Delta M_V)  =
f_\rho(\rho) f_{\Delta \mu} (\Delta \mu) f_{\Delta MV}(\Delta M_V)$.

Each companion can be either physical or optical. So the probability
of physical association can be estimated as
\begin{equation}
P_{\rm phys}(\rho, \Delta \mu, \Delta M_V) = f_{\rm phys}/(f_{\rm phys} + f_{\rm
  opt}) 
\label{eq:phys}
\end{equation}
provided that  the distributions $f_{\rm phys}$ and  $f_{\rm opt}$ are
normalized in the same way. We list estimates of $P_{\rm phys}$ in the
catalog, but  do not  use them in  the statistical  analysis presented
below.   The values  of  $P_{\rm  phys}$ cluster  near  zero and  one,
discriminating  well  most   candidates.   Only  133  companions  have
intermediate $P_{\rm  phys}$ between  0.2 and 0.8,  most of  them with
separations of  few hundred arcseconds.   We checked the  estimates of
$P_{\rm  phys}$ on  175  companions with  trigonometric parallaxes  of
$\pm5$\,mas  or better  precision and  found that  for all  pairs with
$P_{\rm phys}>0.5$ the parallaxes  of primary and secondary companions
match  within errors,  thus confirming  the physical  nature  of those
pairs. However, 39 wide pairs with $P_{\rm phys}<0.5$ (out of 70) also
have  parallaxes  that match  within  5\,mas  and  are probably physical. Our
estimates  of $P_{\rm  phys}$  based on  empirical  data modeling  are
therefore useful as a guidance but should not be taken too literally.

\citet{LB07} also included the mean proper motion of the pair $\mu$ in
their statistical model and showed that the  probability  of  optical
companions is proportional to $\mu^{-3.8}$, i.e. that CPM pairs with
larger proper motions are more likely to be physical. This analysis,
however, neglected the role of color-magnitude selection in
eliminating background contaminants, here represented by $f_{\Delta
  MV}$. In addition, LB2007 used $\Delta \mu \propto  \rho^{-1}$ as a
cutoff, which  introduced incompleteness at  large separations by
restricting the range of $\Delta \mu$ values. For these reasons we
decided not  to include  $\mu$ in  the statistical  criteria used  to
separate physical and optical  companions, instead relying on $f_{\Delta
  MV}$ to eliminate most optical pairs at large angular separations.

To summarize, the current catalog of CPM companions to {\it Hipparcos} stars
contains a  mixture of physical  and optical companions with different
statistical properties such as distributions of separations, $\Delta
\mu$, magnitudes, and colors. We develop a model for these
distributions which we use to evaluate the probability of membership
in either of the two groups. Additional parameters, such as total
proper motion $\mu$, could be used to refine those probability
estimates, but are not applied here for simplicity and uniformity.

\section{Statistics of wide binaries}
\label{sec:stat}

\subsection{Masses and mass ratios}
\label{sec:q}

The masses of the primary and secondary components  are evaluated  here from
their  absolute magnitudes  $M_K$  using the  relation  from \citet{HM93}
 for  dwarfs  and assuming  that  both companions  are
single (although  binarity has little effect on  estimated mass).  The
mass-magnitude relation in  the $K$ band is less  sensitive to age and
metallicity,  compared to  optical magnitudes.   Evolved stars do not
obey  the relation for  dwarfs, and as a result are assigned masses
above  $1.5\,M_\odot$ owing to  their high  luminosity. In any case,
87\% of primary  components  have estimated  masses  between 0.6  and
1.5\,$M_\odot$,  i.e. are nearby solar-type  dwarfs. The narrow range
of masses for the primaries is of course achieved by design, from a
combination of the 67\,pc distance limit and the magnitude limit of the
{\it Hipparcos} catalog itself. Massive stars are rare in the vicinity of
the Sun, and most $<$0.6\,$M_\odot$ dwarfs in the 67\,pc volume are
below the {\it Hipparcos} magnitude limit. The secondary components,
in contrast, span a range of masses from 0.1\,$M_\odot$ up to the mass
of the primary component, consistent with our broad detection of
secondary stars down to the $V=20$ magnitude limit of the SUPERBLINK
survey.

\begin{figure}[t]
\epsscale{1.1}
\plotone{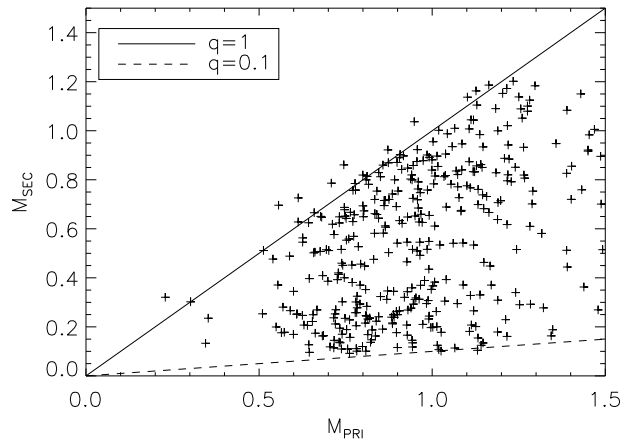}
\caption{Masses of primary $M_{\rm PRI}$ and secondary $M_{\rm SEC}$ components of
  403 systems  with $P_{\rm phys} >  0.5$.  The full  and dashed lines
  indicate mass ratios of 1 and 0.1, respectively.
 \label{fig:massplot}}
\end{figure}

A comparison of the estimated masses of the primary and secondary
components is shown in Figure~\ref{fig:massplot} for the 403 pairs
with $P_{\rm phys} > 0.5$. The majority  of mass  ratios  $q =
M_{\rm SEC}/M_{\rm PRI}$ are  comprised between 0.1 and 1; we set $q=1$ in a few cases
where $M_{\rm SEC} > M_{\rm PRI}$ (which happens because the $V$ magnitudes are used to
define which star is the primary/secondary whereas the $K$ magnitudes
are used to estimate their masses).

The scarcity  of very low-mass companions  is obvious in  the CMD
(Figure~\ref{fig:cmd}) by the scarcity of points with $V - J > 5.5$ or
spectral types  later than M6V.   At 67\,pc (distance modulus  4.12) a
$V=19$  star near  the  SUPERBLINK magnitude  limit  corresponds to  a
0.12\,$M_\odot$  dwarf with  $V-J =  5.2$, according  to  the standard
relations  of \citet{HM93}.   Therefore, the  catalog is incomplete
below $M_{\rm SEC} \sim 0.12\,M_\odot$,  although CPM companions of lower mass
are found around some of the nearest primaries in the catalog.

\begin{figure}[t]
\epsscale{1.1}
\plotone{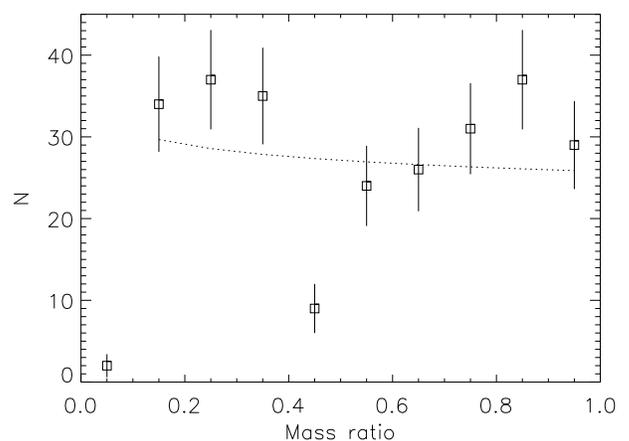}
\caption{Distribution of  the mass ratio  $q = M_{\rm SEC}/M_{\rm PRI}$ for  275 pairs
  with $0.7 <  M_{\rm PRI} < 1.2\,M_\odot$ and $P_{\rm  phys}>0.5$. The dotted
  line shows a power-law fit $f(q) \propto q^{-0.07}$.
 \label{fig:qhist}}
\end{figure}

The mass ratio of wide binaries has approximately uniform distribution
(Figure~\ref{fig:qhist}).   Excluding   the  first  bin   affected  by
incompleteness, we fit  the distribution by a power  law $f(q) \propto
q^{-\beta}$  with  $\beta  =   0.07$.   This  matches  the  result  of
\citet{Raghavan10} who found uniformly  distributed mass ratio for all
binaries   independently  of   their   period,  and   the  result   of
\citet{Tok11}  who studied the  mass-ratio distribution  of companions
within $20\arcsec$ and also found it to be uniform.  Hierarchical multiples
are neglected by this analysis.

However, we  observe a  prominent local minimum  around $q  \sim 0.5$,
which  is  apparent   in  Figure~\ref{fig:qhist}  and  corresponds  to
companions of $\sim  0.5 M_\odot$ mass with $[M_V]_{\rm  SEC} \sim 10$
and $V-J \sim  3.4$.  At a typical distance  of 50\,pc such companions
have $V_{\rm SEC} \sim 13.5$.   As noted above, SUPERBLINK is expected
to be complete in this  magnitude range. However, systematic errors in
the  $V_{\rm SEC}$  magnitudes  may  affect the  position  in the  CMD
(Figure~1),  leading to  a  rejection  of  some physical  companions.
Recently \citet{Rica}  found CPM  companions to exo-planet  hosts with
high  proper motion;  one of  those, HIP~112414  ($\rho  = 51\arcsec$,
$V_{\rm SEC}  = 14.1$, $\mu=214$\,mas~yr$^{-1}$) is  overlooked in our
catalog. Yet, we see only  few such potentially rejected companions on
the right side of the  MS band in Figure~\ref{fig:cmd}, which suggests
that  such  rejections should  be  relatively  unfrequent, and  cannot
explain the observed deficit at $q \sim 0.5$.

Companion masses estimated from  the $K_{\rm SEC}$ magnitudes are not affected
by any systematic errors of $V_{\rm SEC}$. However, the standard relation of
\citet{HM93} used here  changes slope at $M_K =  5.9$, contributing to
the deficit of  estimated masses around $\sim 0.5  M_\odot$. Use of an
alternative smooth relation between mass and $M_K$ decreases the depth
of  the minimum  around $q  \approx 0.5$, without however getting rid
of it entirely.

Should the minimum be real, it might indicate a bimodal distribution
in the masses of the wide secondaries. A population of roughly
equal-mass systems, combined with another where the secondaries have a
mass distribution similar to field stars, could produce a bimodal
distribution of mass ratios such as the one suggested here. Because of
the relatively small number of systems in the present study, this
apparent deficit at $q \approx 0.5$ should be confirmed 
before it can be accepted as real.

\subsection{Projected separations}
\label{sec:s}

\begin{figure}[t]
\epsscale{1.1}
\plotone{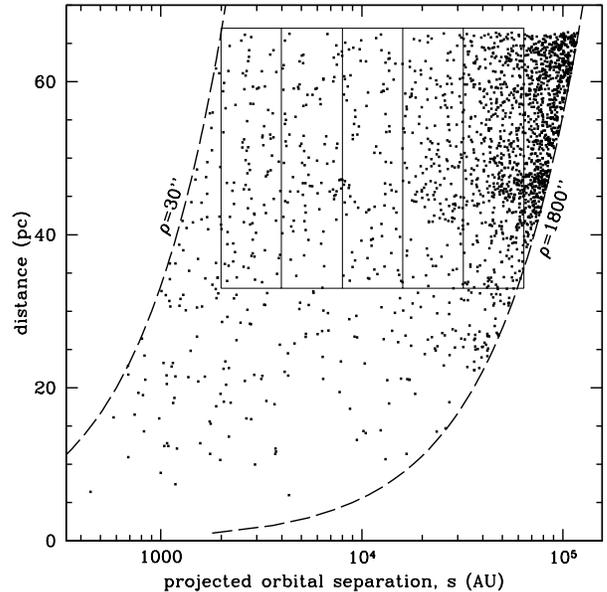}
\caption{Distribution of projected orbital separations plotted against
  distance  from the  Sun.  The projected  separations are  calculated
  assuming the secondary  is at the same distance from  the Sun as the
  Hipparcos  primary.  Angular separation  limits  of  the survey  are
  indicated by  the dashed line.  Our census is complete  for physical
  pairs  with  $s>2\,000$\,AU,  but  contamination from  optical  pairs
  becomes significant  at $s>20\,000$\,AU. The boxes show  the range of
  the five bins in separation and distance used in our estimate of the
  distribution of orbital separations.
 \label{fig:diagram}}
\end{figure}

Projected  orbital  separations  $s$  for  our  wide  systems  can  be
calculated  from  $s  =  \rho/  \pi_{\rm  HIP}$.   The  separations  are
statistically related to the  orbital semi-major axis $a$. Simulations
show that in most cases the ratio $s/a$ is comprised between 0.5 and 2
and the  median of $s/a$ is close  to 1.0 (it depends  slightly on the
eccentricity  distribution).   We  plot  the distribution  of  orbital
separations  against  the  distance  of  the  pair  from  the  Sun  in
Figure~\ref{fig:diagram}. The  range of projected  orbital separations
over which our  survey is sensitive is defined by  the lower and upper
limits  of  angular  separations,   which  are  noted  by  the  dashed
lines. The majority of the  pairs have $33\,{\rm pc} <d<67\,{\rm pc}$;
over  that range  our census  is complete  for systems  with projected
orbital  separations $2\,{\rm  kAU} <s< 64\,{\rm  kAU}$.   In that
regime, the  distribution of projected angular  separations appears to
be  uniform over log($s$)  up to  a projected  separation of  10\,kAU,
beyond which  the density  increases significantly.  This  increase in
density marks  the increasing  contamination from optical  pairs (i.e.
chance alignments).

\begin{figure}[t]
\epsscale{1.1}
\plotone{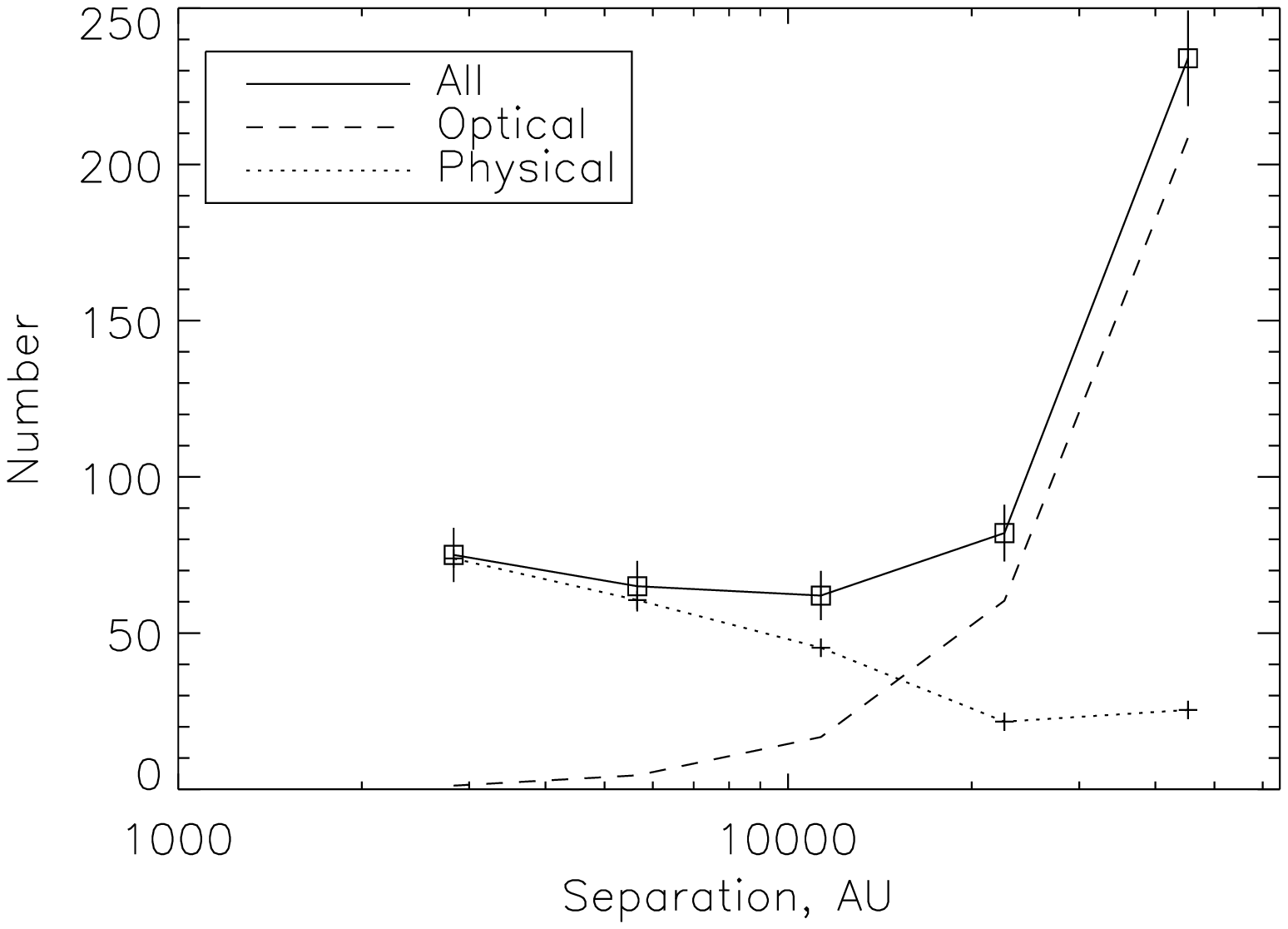}
\plotone{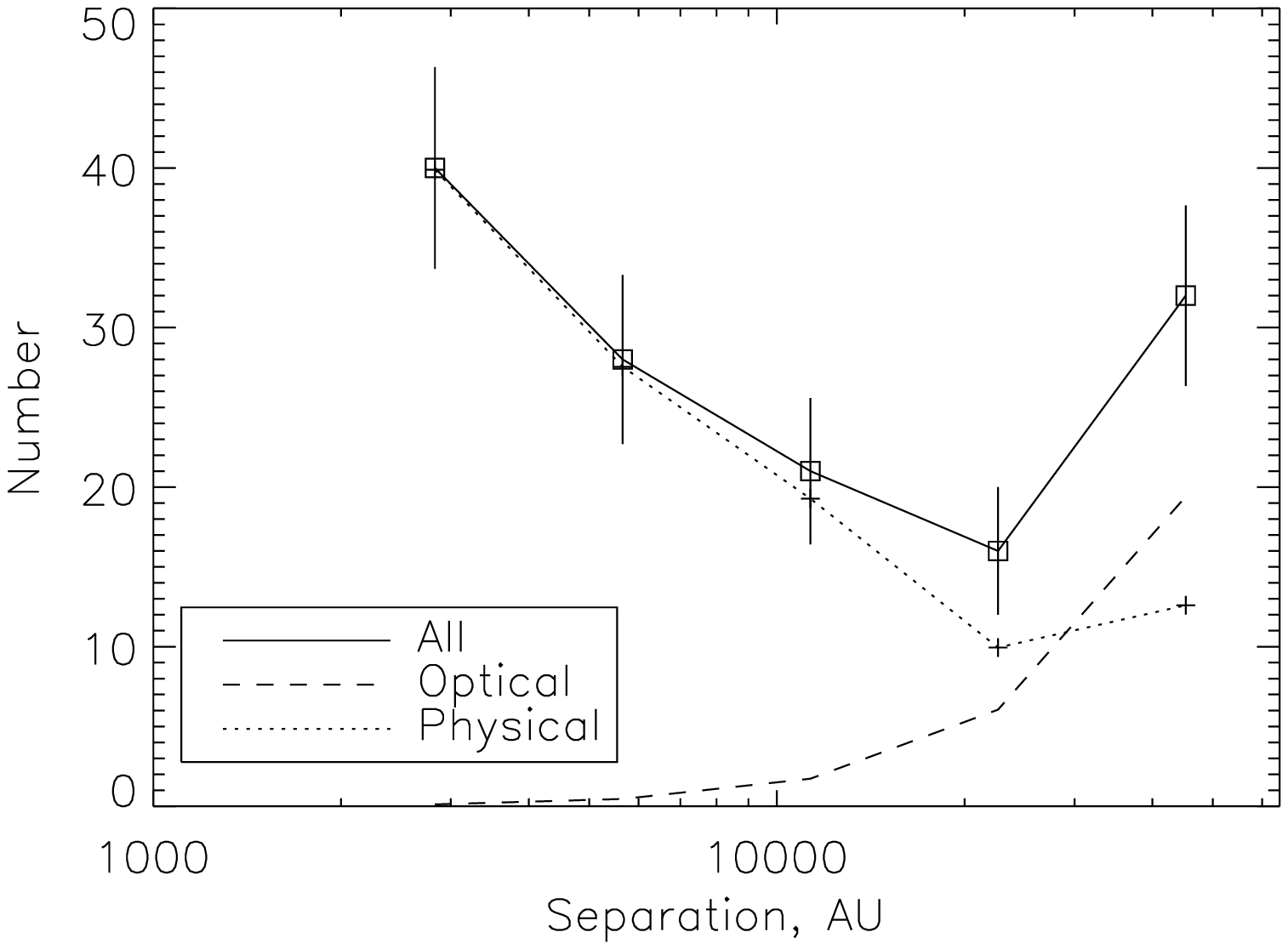}
\caption{Distribution of projected separations for catalog subset with
  primary  masses  from 0.7  to  1.4\,$M_\odot$  and parallax  between
  15\,mas and 30\,mas. Top panel: without restriction on proper motion
  ($N=518$), 
  bottom   panel:  $\mu  >   150$\,mas  yr$^{-1}$,   $N=137$,  $N_{\rm
    ref}=3277$.  Full  line and squares --  observed histogram, dashed
  line  -- estimated  optical companions,  dotted line  and  pluses --
  estimated physical companions.
 \label{fig:rhist}}
\end{figure}

For evaluating  the companion frequency, we define  the {\em reference
  sample} of {\it Hipparcos} stars that fulfill the selection criteria
of  our  catalog  on   distance  ($\pi_{\rm  HIP}  >15$\,mas)  and
proper motion ($\mu>$40\,mas yr$^{-1}$  north of $-20^\circ$  and
$\mu>$150\,mas yr$^{-1}$ otherwise). We  evaluate their  masses by the
same method as  for the catalog targets using $K$-band  magnitudes
from XHIP \citep{XHIP}; for a  small subset  without 2MASS
photometry, $V$  magnitudes  were used instead. The reference sample
is not complete within its distance limit. It only serves to relate  the number of  CPM
companions in  our catalog to  the total number of primary stars with
similar characteristics.

In the following, we  apply additional selection criteria on $\pi_{\rm
  HIP}$, $\mu$, and mass  of primary components to both  our catalog
and reference sample. We also limit the analysis to primary stars in
the mass range 0.7\,$M_\odot<$M$<$1.4\,$M_\odot$, which includes 77\%
of the pairs in our catalog.

To  derive  the distribution  of  projected  separations,  we need  to
account for the  optical companions which contaminate the sample. We
cannot use  here the estimates of $P_{\rm  phys}$ because  they depend
on  $\rho$ and hence  bias the result.  Considering a restricted
range of parallax values, we  estimate the number of  optical
companions in five bins of projected orbital separations (see
Figure~\ref{fig:diagram}), by translating their limits $(s_1,s_2)$ into
angular separation limits $(\rho_1,  \rho_2)$ using the average
parallax of the selected  stars. Then  $N_{\rm   opt}  =  A_{\rm
  opt}(\rho_2^2  - \rho_1^2)$,  where the  parameter $A_{\rm  opt}$ is
evaluated  from the  number  of companions  between 1200\arcsec  and
1800\arcsec, assuming they are all optical in that range (a slight
over-estimate). The estimate of  $ A_{\rm opt}$ relies mostly on
companions with $s > 64$\,kAU, outside the range of separations
studied here.

The  procedure  to  derive  the  separation distribution  is  thus  as
follows. A  subset of the catalog satisfying  some additional criteria
is chosen and the number of reference targets $N_{\rm ref}$ satisfying
the  same criteria is  found. The  histogram of  projected separations
$N_i$  is  constructed  and  the  estimated  number  of  optical  pair
contaminants is subtracted from each bin $i$ to get $N_{i, {\rm phys}}
= N_i  - N_{i,  {\rm opt}}$.  The  companion frequency is  $N_{i, {\rm
    phys}}/N_{\rm  ref}$.  We  ignore the  tiny incompleteness  of the
fifth (32--64)\,kAU bin.

The  top panel of Figure~\ref{fig:rhist} shows the result  for all
stars with solar-type primaries (masses from 0.7\,$M_\odot$ to
1.4\,$M_\odot$). This selects 518 companions, 227 of which  should be
physical, while $N_{\rm ref} = 5196$.  The  total frequency of
companions between  2\,kAU and 64\,kAU is therefore $4.4 \pm
0.3$\%. We do not find any strong dependence of companion fraction on
the mass of the primary. The total binary fraction, however,
critically depends on the subtraction of optical pairs, especially in
the the last two bins which are significantly affected by the
contamination. The first three bins  contain 180 physical
companions, or 3.8$\pm$0.3\% per decade of separation, and are not
much affected by the contamination. 

Our catalog is incomplete for companions of very low masses ($q<0.15$).
The magnitude of this incompleteness depends on the actual
distribution of $q$: if the distribution is uniform all the way down
to $q=0$, then the catalog misses only 15\% of low-mass companions.
However, the  number of sub-stellar  companions is probably  small
(the  so-called brown-dwarf desert)  and the incompleteness  at low
$q$ is  correspondingly less. Wide  triples may also be
under-represented due to our color-magnitude selection (see
\S\ref{sec:mult}), though the losses are probably modest. Assuming
that our census of wide companions is 80\%  complete, their true
frequency would be 5.5\%. This value corroborates the result from
\citet{Raghavan10}, who estimated that 5.9\% of nearby dwarfs have
companions with $s > 2$\.kAU.

More critical to the analysis is whether the distribution of orbital
separations is flat with log $s$ or undergoes a significant
decline. By limiting our analysis to subsets of pairs with larger
proper motions, we can  substantially reduce the optical
contamination, although at  the expense of sample size. Both the
distributions of orbital separations $s$ and the total companion
fraction derived from this restricted sample (bottom panel of
Figure~\ref{fig:rhist}) are very similar to the results  for the full
sample, and strongly support a decline with log $s$.  We  see only  a
slight reduction of the companion fraction with increasing $\mu$
cutoff: $3.6 \pm 0.3$\% for $\mu > 100$\,mas~yr$^{-1}$ and $3.3 \pm
0.3$\% for $\mu > 150$\,mas~yr$^{-1}$. The slight reduction for stars
with higher proper motions could be consistent with those stars being
statistically older, and thus more susceptible to disruption. 
In addition, the distribution in  $\log \rho$ is also declining, as
evidenced by $b=-0.29$ in  the model  (\ref{eq:model}). The  slope is
$-0.29 \times  1.78 =  -0.52$  per decade  (the  range  of $\rho$  is
1.78\,dex),  in  excellent agreement  with  $l=1.5$  found  here by  a
different method.  

\begin{figure}[t]
\epsscale{1.1}
\plotone{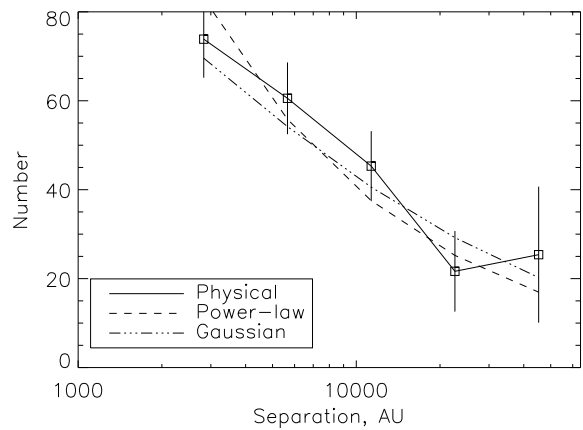}
\caption{Distribution of projected  separations for FGK dwarfs derived
  in this paper (line  and squares) compared to two models consistent
  with the data -- Gaussian \citep{Raghavan10} and a power law with
  $l=1.57$ (dashed line). \label{fig:R10}}
\end{figure}

The distribution  of physical companions  for Solar-mass stars  in the
vicinity of the Sun in $\log s$ is thus declining faster than \"Opik's
law:  $f(s) \propto  s^{-l}$  with  $l \sim  1.5$  (where $l=1$  would
correspond to  \"Opik's law). This  conclusion is valid at  least over
the 2\,kAU$<s<$20\,kAU  range. A deficit  of wide binaries  at large
separations  (or  periods)  relative  to  the \"Opik's  law  was  also
documented by \citet{Chaname05}, and suggested by \citet{LB07} using a
subset of our current data. This decline is supported by the current,
extended sample.  \citet{LB07} found  a declining distribution with $l
\approx  1.6$ at  $s  >  3$\,kAU; the  slightly  faster decline  could
however be  the result of  an additional selection  criterion ($\Delta
\mu \propto \rho^{-1}$) which favors closer binaries.

Figure~\ref{fig:R10}   compares    the   empirical   distribution   of
separations  with two possible analytical  models: a simple power law,
and the log-normal period distribution   of   \citet{Raghavan10}. For
the latter, the median   period $10^5$\,d, logarithmic dispersion
2.28, and binary  frequency 0.5 was converted to the distribution  of
semi-major  axis  by  assuming total  mass  of 1.5\,$M_\odot$  and
total sample  size  $N_{\rm  ref} = 5196$.   The agreement with the
log-normal distribution is better than one might expect, given  the
fact that their sample  does not overlap with  the stars  studied here
and  that the analysis methods are different.

The frequency of 500-AU companions to  FG dwarfs is  higher, 13\% per
decade of separation \citep{Tok11}.  This agrees with the distribution
declining  at long  orbital periods.   Indeed, extrapolation  from the
first three bins (geometric mean  $s = 5.65$\,kAU, frequency 3.8\% per
decade) to $s \sim 500$\,AU with the $s^{-1.5}$ power law predicts 3.4
times  more  companions,  13\%   per  decade.   In  other  words,  the
$s^{-1.5}$ law seems to hold at separations less than 1\,kAU.

\section{Summary and discussion}
\label{sec:sum}

We have searched for wide companions to {\it Hipparcos} stars in the
SUPERBLINK proper motion catalog based on common proper motion. The
main results are:
\begin{itemize}
\item
The mass ratio of wide companions to nearby solar-mass dwarfs is found
to be distributed uniformly to first order, with a possible deficit of
companions with mass ratios $q\simeq0.5$. 
\item
The distribution of projected  separation $s$ declines as $f(s) \propto
s^{-1.5}$, faster than the \"Opik's  law. 
\item
Total  frequency   of  wide  companions  to   solar-type  dwarfs  with
$2<s<64$\,kAU is $4.4  \pm 0.3$\%.  The frequency in the  range from 2 to
16\,kAU is $3.8 \pm 0.3$\%  per decade of separation. Both numbers are
lower limits because a fraction of wide companions is missed.
\end{itemize}
In addition, our search has produced a catalog of 1413 CPM companions
to {\it Hipparcos} stars.
The  present  catalog gives  the  most  complete  census of  wide  CPM
companions to nearby  stars available to date. A  decline in companion
frequency  at  large   $s$  found  here  can  be   used  to  constrain
gravitational perturbations in the Galactic field.

We   found    21   potential    WD   companions   to    nearby   stars
(Table~\ref{tab:wd}). Half  of those should  be true WDs, the  rest are
faint background stars.  Indeed, 5 WD companions are already known in
the literature, 6 look promising based on their separation and PM. 

Our  catalog contains  new wide  companions to  exo-planet  host stars
\citep{Schneider},  in addition  to previously  known  CPM companions.
Relevant data on two high-probability physical companions to HIP~17747
and HIP~90593  are listed in  Table~\ref{tab:exoplanet}, the potential
WD  companion  to HIP~60081  is  listed  in Table~\ref{tab:wd}.   More
information  on these candidates  (spectral types,  radial velocities)
are necessary for their definitive attribution.






\acknowledgments  This material is based upon work supported by the
National Science Foundation under Grants No. AST 06-07757, AST
09-08419. This work used  the SIMBAD service operated  by  Centre des
Donn\'ees  Stellaires (Strasbourg,  France), bibliographic references
from  the Astrophysics Data System maintained by SAO/NASA,  data
products of  the Two Micron All-Sky  Survey (2MASS) and the Washington
Double Star Catalog maintained at USNO. We thank the referee,
J.~Subasavage, for  thorough and timely examination of our work.  
%


\clearpage
\begin{landscape}


\begin{deluxetable}{r r rr rr ccccc c c l c c c ccc}
\tabletypesize{\scriptsize} 
\tablewidth{0pt}
\tablecaption{Catalog of wide CPM companions. I. Primaries}
\tablehead{
\colhead{HIP$_{\rm PRI}$} & 
\colhead{PM Id} & 
\colhead{R.A.} & 
\colhead{Dec.} & 
\colhead{$\mu_\alpha$} & 
\colhead{$\mu_\delta$} & 
\colhead{$V$} & 
\colhead{$V-J$} & 
\colhead{$J$} & 
\colhead{$H$} & 
\colhead{$K$} & 
\colhead{$\pi$} & 
\colhead{$\sigma_\pi$} & 
\colhead{code} &
\colhead{Sep} & 
\colhead{$\Delta \mu$} & 
\colhead{$P_{\rm phys}$} \\
   & & \colhead{(deg)} & 
\colhead{(deg)} & 
(as yr$^{-1}$) &  (as yr$^{-1}$) & (mag) & (mag) &  (mag) & (mag) & (mag) & (mas) & (mas) &  & (arcsec) &   (mas yr$^{-1}$) &  \\ 
\colhead{(1)} &
\colhead{(2)} &
\colhead{(3)} &
\colhead{(4)} &
\colhead{(5)} &
\colhead{(6)} &
\colhead{(7)} &
\colhead{(8)} &
\colhead{(9)} &
\colhead{(10)} &
\colhead{(11)} &
\colhead{(12)} &
\colhead{(13)} &
\colhead{(14)} &
\colhead{(15)} &
\colhead{(16)} &
\colhead{(17)} 
}
\startdata
   159 & I00020-0245 &    0.510909 &   -2.766169 &  0.040 & -0.003 &   6.96 &   0.63 &   6.33 &   6.27 &   6.21 &   16.0 &    0.5 & PLX1 & 1715.8 &   11.2 &  0.003\\
   238 & I00029-2002 &    0.739543 &  -20.045879 &  0.103 &  0.081 &   6.30 &   1.02 &   5.28 &   5.07 &   4.96 &   16.3 &    0.5 & PLX1 & 1207.2 &   20.2 &  0.025\\
   277 & I00034-3615 &    0.859494 &  -36.251194 &  0.071 &  0.005 &   7.00 &   0.67 &   6.33 &   6.23 &   6.15 &   15.2 &    0.5 & PLX1 & 1470.3 &   17.0 &  0.006\\
   473 & I00056+4548 &    1.420712 &   45.812103 &  0.879 & -0.154 &   8.20 &   2.06 &   6.14 &   4.71 &   5.28 &   88.4 &    1.6 & PLX1 &  328.2 &    9.5 &  0.723\\
   493 & I00059+1814 &    1.478141 &   18.235023 & -0.151 & -0.150 &   7.45 &   1.12 &   6.33 &   6.08 &   6.03 &   26.9 &    0.6 & PLX1 &  573.0 &    3.2 &  0.741\\
   577 & I00070+3252 &    1.752726 &   32.870110 &  0.177 & -0.077 &   9.61 &   1.39 &   8.22 &   7.86 &   7.72 &   16.1 &    1.3 & PLX1 & 1401.7 &    5.0 &  0.024\\
   599 & I00072-4153 &    1.814071 &  -41.889565 & -0.063 &  0.018 &   8.29 &   1.21 &   7.08 &   6.82 &   6.72 &   16.9 &    0.8 & PLX1 &  653.1 &   16.1 &  0.021\\
   682 & I00084+0637 &    2.107265 &    6.616799 &  0.085 & -0.003 &   7.66 &   1.24 &   6.42 &   6.15 &   6.12 &   25.6 &    0.7 & PLX1 & 1776.1 &   11.7 &  0.000\\
   731 & I00090+2739 &    2.262037 &   27.651575 &  0.223 &  0.147 &  11.65 &   2.23 &   9.42 &   8.78 &   8.66 &   23.5 &    2.9 & PLX1 &   68.8 &   11.7 &  0.927\\
   840 & I00103-0514 &    2.578626 &   -5.248588 &  0.037 & -0.029 &   5.95 &   1.94 &   4.01 &   3.38 &   3.37 &   17.1 &    0.5 & PLX1 & 1561.7 &   12.1 &  0.017\\
\enddata
\end{deluxetable}

\setcounter{table}{0}
\begin{deluxetable}{r r r rr rr ccccc c c l }
\tabletypesize{\scriptsize} 
\tablewidth{0pt}
\tablecaption{Catalog of wide CPM companions. II. Secondaries}
\tablehead{
\colhead{HIP$_{\rm PRI}$} & 
\colhead{HIP$_{\rm SEC}$} & 
\colhead{PM Id} & 
\colhead{RA(2000)} & 
\colhead{DEC(2000)} & 
\colhead{$\mu_\alpha$} & 
\colhead{$\mu_\delta$} & 
\colhead{$V$} & 
\colhead{$V-J$} & 
\colhead{$J$} & 
\colhead{$H$} & 
\colhead{$K$} & 
\colhead{$\pi$} & 
\colhead{$\sigma_\pi$} & 
\colhead{code} \\
&   & & \colhead{(deg)} & 
\colhead{(deg)} & 
(as yr$^{-1}$) &  (as yr$^{-1}$) & (mag) & (mag) &  (mag) & (mag) & (mag) & (mas) & (mas) &  \\
\colhead{(1)} &
\colhead{(18)} &
\colhead{(19)} &
\colhead{(20)} &
\colhead{(21)} &
\colhead{(22)} &
\colhead{(23)} &
\colhead{(24)} &
\colhead{(25)} &
\colhead{(26)} &
\colhead{(27)} &
\colhead{(28)} &
\colhead{(29)} &
\colhead{(30)} &
\colhead{(31)} 
}
\startdata
   159 & 999999 & I00002-0256 &    0.069661 &   -2.947790 &  0.045 & -0.013 &  10.00 &   1.36 &   8.64 &   8.37 &   8.23 &   11.9 &    0.0 & PHOT \\
   238 & 999999 & I00032-1943 &    0.808330 &  -19.716831 &  0.086 &  0.070 &  15.46 &   4.09 &  11.37 &  10.79 &  10.52 &   18.4 &    0.0 & PHOT \\
   277 & 999999 & I00034-3550 &    0.868067 &  -35.842834 &  0.079 &  0.020 &   9.61 &   1.77 &   7.84 &   7.37 &   7.22 &   23.7 &    0.0 & PHOT \\
   473 &    428 & I00051+4547 &    1.295195 &   45.786587 &  0.870 & -0.151 &   9.95 &   3.25 &   6.70 &   6.10 &   5.85 &   88.9 &    1.4 & PLX1 \\
   493 &    495 & I00059+1804 &    1.481153 &   18.075890 & -0.150 & -0.147 &   8.58 &   1.46 &   7.12 &   6.72 &   6.46 &   26.9 &    1.2 & PLX1 \\
   577 & 999999 & I00061+3312 &    1.534962 &   33.214226 &  0.173 & -0.074 &  16.16 &   3.99 &  12.17 &  11.63 &  11.37 &   11.9 &    0.0 & PHOT \\
   599 & 999999 & I00082-4154 &    2.055929 &  -41.911892 & -0.049 &  0.010 &  11.76 &   2.00 &   9.76 &   9.18 &   9.06 &   11.0 &    0.0 & PHOT \\
   682 & 999999 & I00065+0645 &    1.630693 &    6.756144 &  0.096 & -0.007 &  12.38 &   2.56 &   9.82 &   9.22 &   9.06 &   14.1 &    0.0 & PHOT \\
   731 & 999999 & I00089+2739 &    2.248174 &   27.666235 &  0.212 &  0.151 &  13.70 &   3.27 &  10.43 &   9.81 &   9.60 &   16.1 &    0.0 & PHOT \\
   840 & 999999 & I00120-0513 &    3.013638 &   -5.225371 &  0.048 & -0.024 &  14.82 &   3.82 &  11.00 &  10.41 &  10.16 &   18.2 &    0.0 & PHOT \\
\enddata
\end{deluxetable}

\clearpage
\end{landscape}

\begin{deluxetable}{r r r   r r r c  c c l }
\tabletypesize{\scriptsize} 
\tablewidth{0pt}
\tablecaption{Candidate White Dwarf CPM Companions
\label{tab:wd} }
\tablehead{
HIP$_{\rm PRI}$ & R.A. & Dec. & Sep & $\Delta \mu$ & $\mu$ & $[\pi_{\rm HIP}]_{\rm PRI}$ & $V_{\rm
  SEC}$ & $(V-J)_{\rm SEC}$ & Comment \\
     & $h\;  m \;  s$ & $^\circ \; ' \; ''$ & (as) & \multicolumn{2}{c}{ (mas yr$^{-1}$) }
  & (mas) & (mag) & (mag) & \\
(1) & (2) & (3) & (4) & (5) & (6) & (7) & (8) & (9) & (10)  
}
\startdata
  8588 &  01 50 52.1 & +11 05 34.1 &  178.0 &  16 &  74 &   23.8 &  20.29 &   2.11  &   \\
 11028 &  02 21 57.9 & +04 45 17.9 &   53.1 &  17 &  82 &   24.7 &  16.67 &   0.81  & wd?  \\
 42783 &  08 42 57.6 & +24 09 30.8 &  104.6 &  23 & 113 &   27.1 &  17.17 &   1.18  &   \\
 50760 &  10 21 43.4 & +59 34 35.1 &  126.5 &   6 &  57 &   19.0 &  19.69 &   2.10  &   \\
 54530 &11 09 30.1 & $-$26 01 06.9 &  100.3 &   5 & 238 &   24.9 &  17.33 &   0.94  & WD1107$-$257  \\
 56004 &11 28 34.9 & $-$45 05 13.8 &  195.3 &  21 &  43 &   15.6 &  18.60 &   2.13  &   \\
 60081 &12 19 13.7 & $-$03 19 46.1 &   35.0 &  19 &  72 &   19.5 &  16.82 &   0.17  & wd?  \\
 63181 &  12 56 33.7 & +36 50 05.6 &  129.8 &  16 &  62 &   24.7 &  19.73 &   2.29  &   \\
 69178 &  14 09 43.0 & +15 39 39.8 &  164.8 &  14 &  73 &   17.8 &  20.45 &   2.31  &   \\
 70446 &  14 24 23.6 & +47 51 02.6 &  177.5 &  15 &  65 &   30.9 &  17.76 &   2.45  &   \\
 80182 &  16 22 03.9 & +12 13 33.6 &   62.4 &  10 &  94 &   19.3 &  15.05 &   0.06  & WD1619+123  \\
 85799 &  17 31 46.7 & +16 51 45.7 &  174.9 &  21 &  41 &   21.2 &  18.04 &   1.74  &   \\
 90593 &  18 28 59.9 & +11 41 07.5 &  167.0 &  24 &  50 &   15.9 &  18.55 &   2.00  &   \\
 99956 &20 16 55.6 & $-$03 27 51.0 &  127.3 &   8 & 108 &   15.4 &  16.55 &   0.12  & wd?  \\
101268 &  20 31 36.4 & +34 21 14.9 &   85.6 &  18 &  53 &   21.8 &  17.57 &   1.82  &   \\
102488 &  20 46 05.3 & +33 58 12.2 &   92.0 &   2 & 485 &   44.9 &  14.68 &   0.64  & GJ 9707C  \\
103735 &21 01 26.7 & $-$35 09 33.4 &  186.0 &   6 & 188 &   21.5 &  16.96 &   0.70  & wd?  \\
106335 &  21 32 16.2 & +00 15 14.4 &  133.1 &  10 & 415 &   22.1 &  14.65 &$-$0.24  & WD2129+004  \\
110218 &  22 19 28.5 & +21 22 19.5 &   83.1 &  13 & 422 &   20.3 &  18.32 &   1.87  & WD2217+211  \\
117011 &23 43 00.8 & $-$64 47 37.8 &   41.6 &  16 & 241 &   27.2 &  17.15 &   1.10  & wd?  \\
118010 &  23 56 09.2 & +59 47 43.6 &  101.8 &   9 & 342 &   19.7 &  18.68 &   1.83  & wd?  
\enddata
\end{deluxetable}



\begin{deluxetable}{l l |c  cc  cc  c c}
\tabletypesize{\scriptsize} 
\tablewidth{0pt}
\tablecaption{New wide companions to exo-planet hosts
\label{tab:exoplanet} }
\tablehead{
HIP$_{\rm PRI}$ & HD$_{\rm PRI}$ &$P_{\rm pl}$ & $\rho$ & $s$ & $\Delta \mu$ & $\mu$ & $P_{\rm  phys}$ & $\Delta M_V$ \\   
    &   & (d)         &  (as) & (kAU)  & (mas yr$^{-1}$)       & (mas
yr$^{-1}$)  &   & (mag)   
}
\startdata
17747 & 23596 & 1656      & 70.7 & 3.6  &  4           &  58 & 0.996 & $-$0.8 \\
90593 & 170469& 1145      & 43.2 & 2.7  &  9           & 51  & 0.989 & $-$1.1 
\enddata
\end{deluxetable}


\end{document}